\documentclass{elsart}
\usepackage{graphicx}
\usepackage{amssymb,amsmath}

\begin{document}
\begin{frontmatter}
\date{22. September 2006}
\vspace{-1.2 cm}
\title{Fast Photon Detection for Particle Identification with COMPASS RICH-1}
\vspace{-0.6 cm}
\small
P.Abbon$^{k}$, M.Alekseev$^{l}$, H.Angerer$^{i}$, M.Apollonio$^{m}$, 
R.Birsa$^{m}$, P.Bordalo$^{g}$, F.Bradamante$^{m}$, A.Bressan$^{m}$, L.Busso$^{l}$, 
M.Chiosso$^{l}$, P.Ciliberti$^{m}$, M.L.Colantoni$^{a}$, S.Costa$^{l}$, 
S.Dalla Torre$^{m}$, T.Dafni$^{k}$, E.Delagnes$^{k}$, H.Deschamps$^{k}$, V.Diaz$^{m}$, N.Dibiase$^{l}$, V.Duic$^{m}$, 
W.Eyrich$^{d}$, 
D.Faso$^{l}$, A.Ferrero$^{l}$, M.Finger$^{j}$, M.Finger Jr$^{j}$, H.Fischer$^{e}$, 
S.Gerassimov$^{i}$, M.Giorgi$^{m}$, B.Gobbo$^{m}$, 
R.Hagemann$^{e}$, D.von~Harrach$^{h}$, F.H.Heinsius$^{e}$, S.Horikawa$^{c}$, 
R.Joosten$^{b}$, 
B.Ketzer$^{i}$, K.K\"onigsmann$^{e}$, V.N.Kolosov$^{c,}$\footnote[1]{\rm On leave from IHEP Protvino, Russia.}, 
I.Konorov$^{i}$, D.Kramer$^{f}$, F.Kunne$^{k}$, 
A.Lehmann$^{d}$, S.Levorato$^{m}$, 
A.Maggiora$^{l}$, A.Magnon$^{k}, $ A.Mann$^{i}$, A.Martin$^{m}$, G.Menon$^{m}$, A.Mutter$^{e}$, 
O.N\"ahle$^{b}$, F.Nerling$^{e,}$\footnote[2]{\rm Corresponding author, E-mail address: \texttt{frank.nerling@cern.ch}~~(F.
Nerling).}
, D.Neyret$^{k}$, 
P.Pagano$^{m}$, S.Panebianco$^{k}$, D.Panzieri$^{a}$, S.Paul$^{i}$, G.Pesaro$^{m}$, J. Polak$^{f}$, 
P.Rebourgeard$^{k}$, F.Robinet$^{k}$, E.Rocco$^{m}$,
P.Schiavon$^{m}$, C.Schill$^{e}$, W.Schr\"oder$^{d}$, L.Silva$^{g}$, M.Slunecka$^{j}$, F.Sozzi$^{m}$,
L.Steiger$^{j}$, M.Sulc$^{f}$, M.Svec$^{f}$, 
F.Tessarotto$^{m}$, A.Teufel$^{d}$, H.Wollny$^{e}$ 
\vspace{-0.3cm}
\address{
\it (a) INFN, Sezione di Torino and University of East Piemonte, Alessandria, Italy \\
\it (b) Universit\"at Bonn, Helmholtz-Institut f\"ur Strahlen- und Kernphysik, Bonn, Germany \\
\it (c) CERN, European Organization for Nuclear Research, Geneva, Switzerland \\
\it (d) Universit\"at Erlangen-N\"urnberg, Physikalisches Institut, Erlangen, Germany \\
\it (e) Universit\"at Freiburg, Physikalisches Institut, Freiburg, Germany \\
\it (f) Technical University of Liberec, Liberec, Czech Republic \\
\it (g) LIP, Lisbon, Portugal \\
\it (h) Universit\"at Mainz, Institut f\"ur Kernphysik, Mainz, Germany \\
\it (i) Technische Universit\"at M\"unchen, Physik Department, Garching, Germany \\
\it (j) Charles University, Prague, Czech Republic and JINR, Dubna, Russia \\
\it (k) CEA Saclay, DSM/DAPNIA, Gif-sur-Yvette, France \\
\it (l) INFN, Sezione di Torino and University of Torino, Torino, Italy \\
\it (m) INFN, Sezione di Trieste and University of Trieste, Trieste, Italy \\
}
\vspace{-0.9cm}
\begin{abstract}
Particle identification at high rates is an important challenge for many 
current and future high-energy physics experiments. 
The upgrade of the COMPASS RICH-1 detector requires a new technique for Cherenkov photon detection at count
rates of several $10^6$ per channel in the central detector region, and a read-out system allowing for trigger
rates of up to 100~kHz. To cope with these requirements, the photon detectors in the central region have been replaced 
with the detection
system described in this paper. In the peripheral regions, the existing multi-wire proportional chambers with 
CsI photocathode are now read out via a new system employing APV pre-amplifiers and flash ADC chips. 
The new detection system consists of multi-anode photomultiplier tubes (MAPMT) and fast read-out 
electronics based on the MAD4 discriminator and the F1-TDC chip. 
The RICH-1 is in operation in its upgraded version for the 2006 CERN SPS run. 
We present the photon detection design, constructive aspects and the first Cherenkov light in the detector.
\end{abstract}
\end{frontmatter}
\vspace{-1.2 cm}
%
%
\section{Introduction \& motivation of the project}
\label{intro}
\vspace{-0.8 cm}
\begin{figure}[t]
\begin{minipage}[c]{.31\linewidth}
\begin{center}
\includegraphics[clip,bb= 75 44 406 582,width=1.\linewidth]{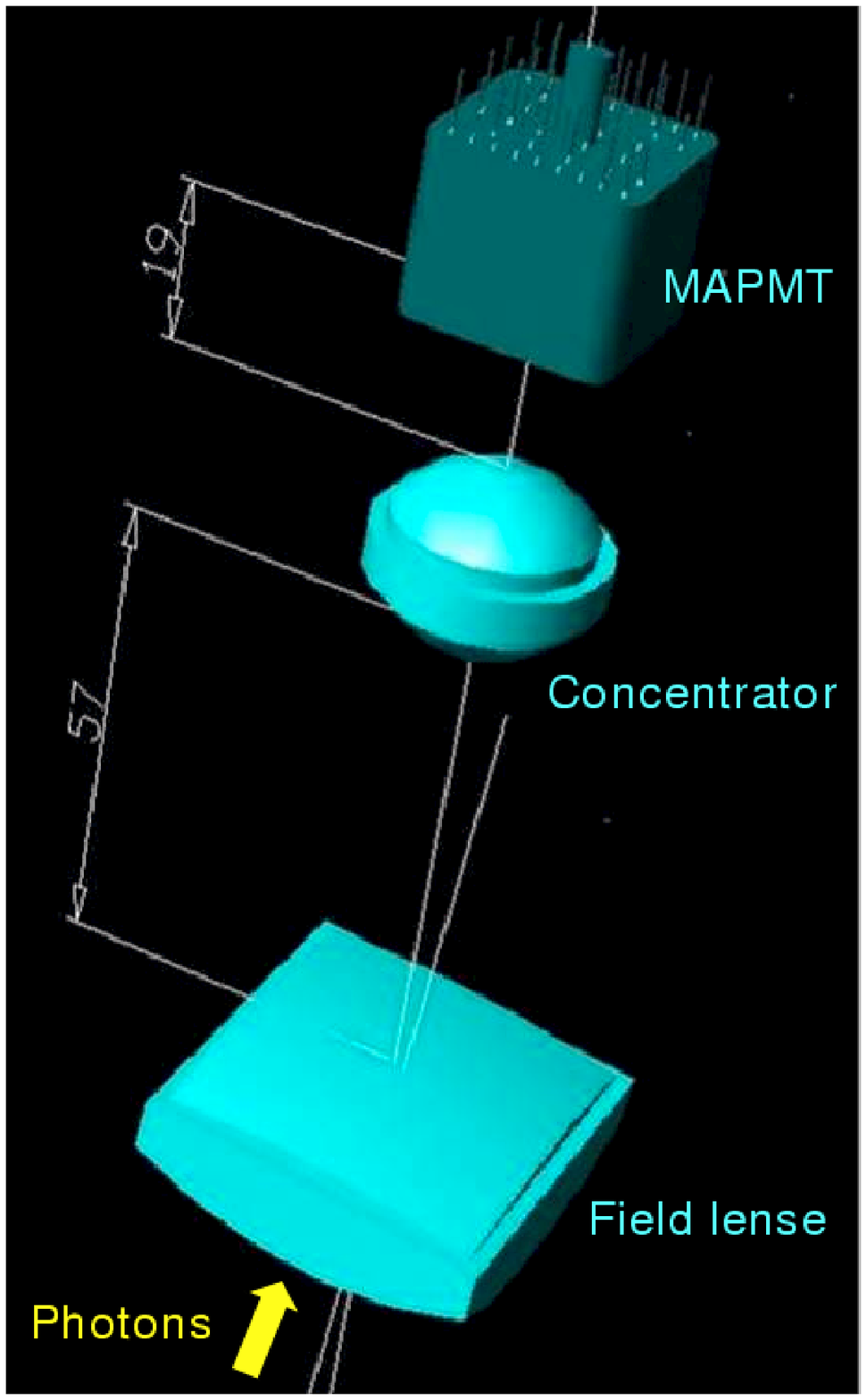}
\end{center}
\caption{Scheme of fused silica telescope system connected to each MAPMT.}
\label{fig1.telescope}
\end{minipage}\hfill
\begin{minipage}[c]{.64\linewidth}
\begin{center}
\includegraphics[clip,bb= 75 43 770 505,width=1.\linewidth]{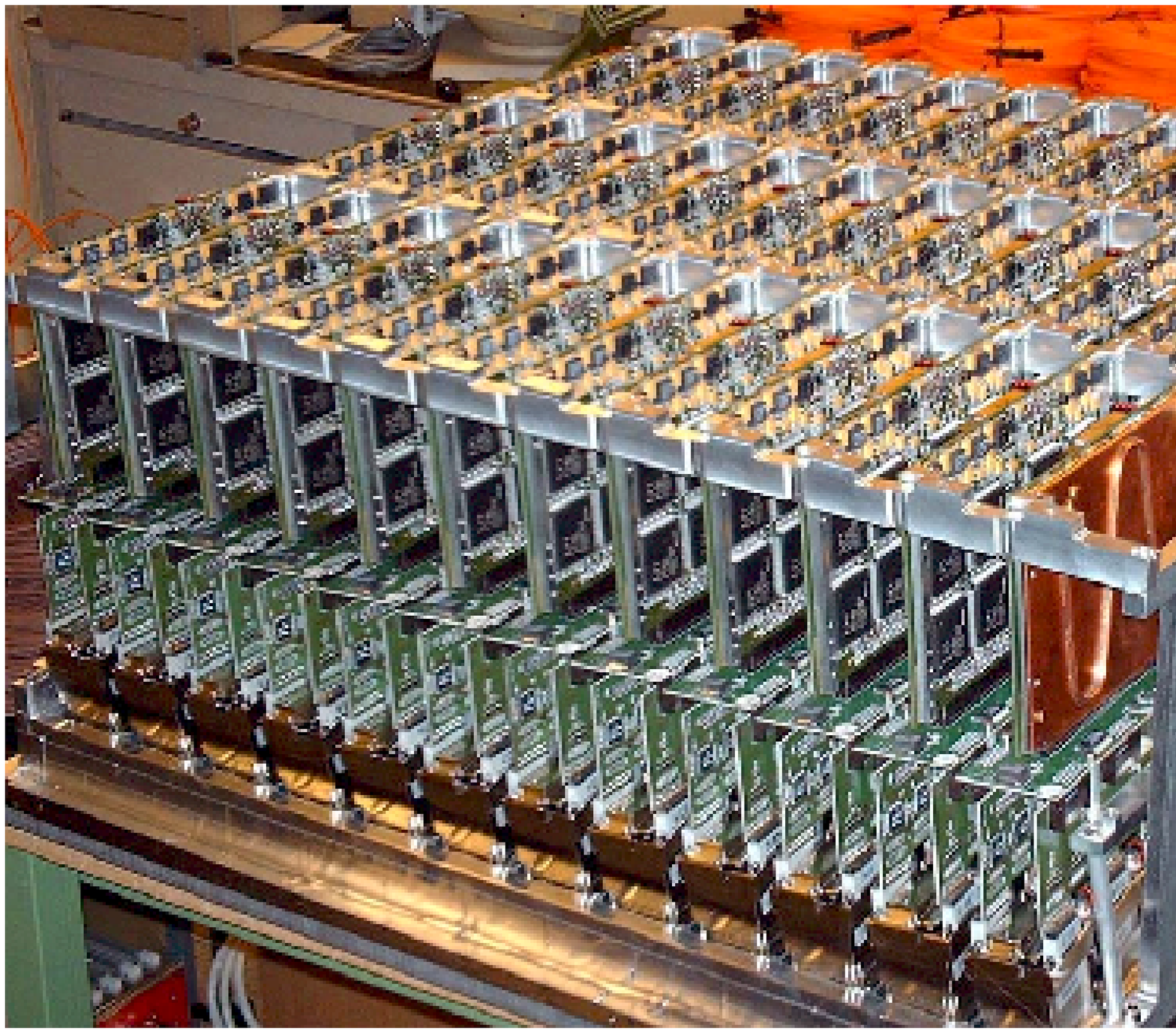}
\end{center}
\vspace{0.1cm}
\caption{First quarter of the new detection system comprising 144 MAPMT, 
fully eqipped with FE electronics. The FE cards are water-cooled by single waterline copper plates, 
as indicated by the three DREISAM boards in the front.}
\label{fig2.quarter}
\end{minipage}
\end{figure}
The fixed target experiment COMPASS \cite{compass} at CERN SPS is a two stage spectrometer dedicated to 
the investigation of perturbative and non-perturbative QCD. The comprehensive research programe comprises both 
physics with a muon and hadron beams, including the study of the nucleon spin structure and charm
spectroscopy. Identification of hadronic particles is required, and performed by RICH-1 
in the multi-decade GeV/c range \cite{albrecht}. For this purpose, a large size gas radiator RICH has been in operation in
COMPASS since 2001. Multi-wire proportional chambers (MWPC) with CsI photocathodes have been used for the
single photon detection, whereas the read-out was based on the Gassiplex front-end (FE) chip \cite{gassiplex}. 
Now, the central quarter of the 5.3\,m$^2$ photon detection area has been fitted with a new detection system
based on MAPMT \cite{mapmt} and is discussed in this paper. The outer regions stay unchanged, and are read out by a new 
system \cite{apv} characterised by negligible dead-time and better time
resolution. Details on the COMPASS data acquisition system may be found in \cite{daq}.
\newline
The memory time of about $3\,\mu$s 
of the detection system and the dominant background due to uncorrelated 
muon beam particles in the central detector region had remarkably reduced the detector 
performance, especially for particles at the very forward direction. Furthermore, the planned increase of the 
beam intensity from 40 to 100\,MHz, and trigger rates from 20\,kHz in the past to 100\,kHz, 
made upgrading the COMPASS RICH-1 mandatory in terms of a faster photon detection system.
\vspace{-0.6 cm}
%
%
\section{Realisation of the project}
\label{realisation}
\vspace{-0.8 cm}
The new detector part consists of 576 MAPMT (Hamamatsu R7600-03-M16) with 16 channels per PMT, 
each coupled to an individual fused silica telescope, see Fig.\,\ref{fig1.telescope}. 
The purpose of the optics is to focus the Cherenkov photons on the sensitive cathodes to gain a factor of 
approximately 7 in sensitive surface. Moreover, the telescope has been designed to minimise image distortions, provide an 
angular acceptance of $\pm 9.5^{\rm o}$, and perform a spot size of $\sim$\,1\,mm (r.m.s). 
The read-out is performed by 4 sensitive MAD4 chips \cite{mad4} on 2 FE cards per MAPMT, and high-resolution F1 TDC \cite{tdc}. 
One so-called DREISAM card housing 8 F1-TDC reads 4 MAPMT. Both electronic cards are water-cooled via copper water line 
plates. One panel of MAPMT fully equipped with FE electronics is shown in Fig.\,\ref{fig2.quarter}. 
The MAD4 chip features a small noise level (5-7\,fC compared to mean signals 
of 500\,fC) and has a rate capability up to $\sim$\,1\,MHz per channel. The new FE chip version CMAD,
that will be available for the 2007 run, will operate up to rates of 5 MHz per channel. 
The F1-TDC operate stably for input rates up to 10\,MHz per channel at 100\,kHz trigger rates, and the time resolution of 120\,ps
further ensures the background level from uncorrelated physics events to be negligible. 
\begin{figure}[t]
\begin{minipage}[c]{.48\linewidth}
\begin{center}
\includegraphics[clip,bb= 175 120 666 576,width=1.\linewidth]{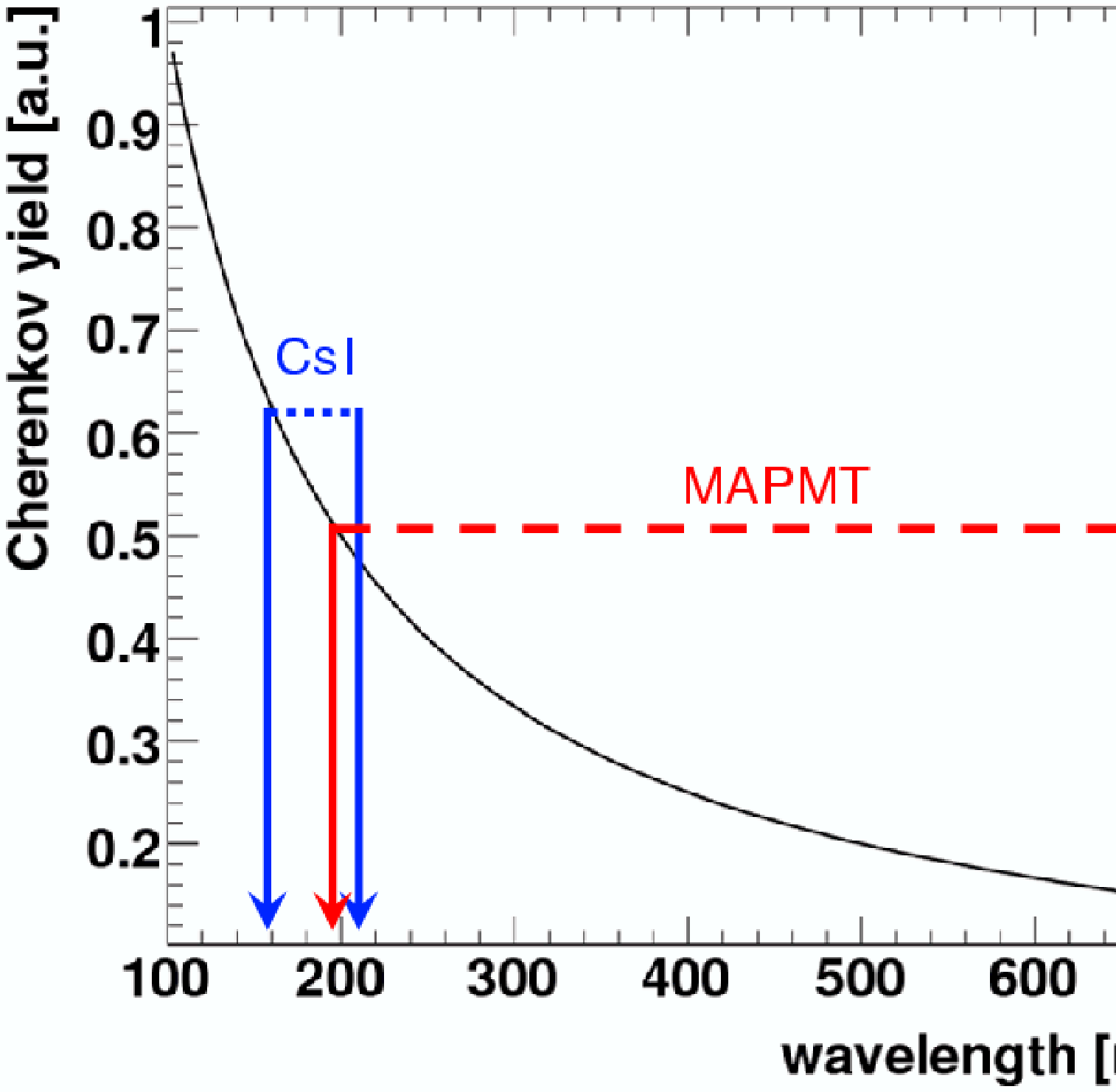}
\end{center}
\caption{Cherenkov yield as a function of wavelength - the different sensitivities for CsI and MAPMT are indicated.}
\label{fig3.yield}
\end{minipage}\hfill
\begin{minipage}[c]{.48\linewidth}
\begin{center}
\includegraphics[clip,bb= 288 260 616 542,width=1.0\linewidth]{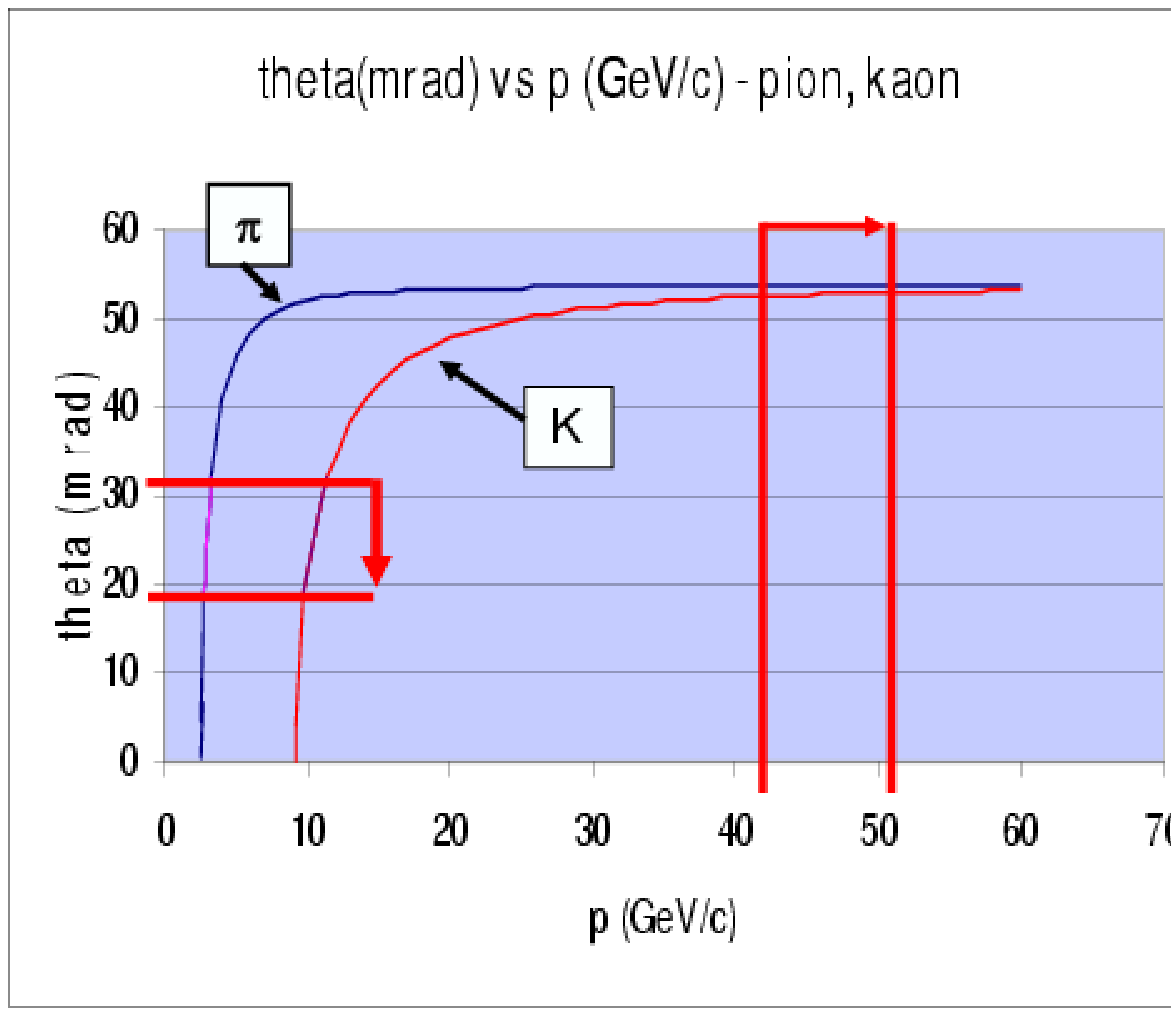}
\end{center}
\caption{Simulated Cherenkov emission angle $\theta_{\rm Ch}$ versus particle momenta - 
the PID capabilities extended by the detector upgrade are marked.}
\label{fig4.PID}
\end{minipage}
\end{figure}
\vspace{-0.6 cm}
%
%
\section{Expected performance \& first light in the upgraded detector}
\label{perform}
\vspace{-0.8 cm}
The performance achieved in the past can be summarised as follows \cite{albrecht}: 
Photons per particle at saturation $N_{{\rm ph} / {\rm ring}} = 14$ (mean value), single photon and global resolution 
on the reconstructed Cherenkov angle $\theta_{\rm Ch}$ at saturation of $\sigma_{\rm ph}=1.2$\,mrad and 
$\sigma_{\rm ring}=0.6$\,mrad respectively, resulting in particle identification
(PID) efficiency better than $95\%$ for $\theta_{\rm Ch}\,>\,30\,\rm{mrad}$, and 2$\sigma$ 
pion-to-kaon separation at 43\,GeV/c. 
\newline
One benefit of using MAPMT is the extension in the wavelength range and the resultant increase of detected Cherenkov 
photons as illustrated in Fig.\,\ref{fig3.yield}. 
We expect an increase in the number of detected photons per ring by approximately a factor of 3 to $N_{{\rm ph} / {\rm ring}}\approx 40$ ($\beta\sim 1$). 
Due to the improved time resolution of a few ns, we further expect a gain in the time resolution of single ring 
reconstruction to $\sigma_{\rm ring} \approx 0.4\,$mrad at $\beta\sim 1$ . 
The improved $N_{{\rm ph} / {\rm ring}}$ leads to an extended PID capability towards $\theta_{\rm Ch} < 30\,\rm{mrad}$ 
and lower particle momenta, whereas the better $\sigma_{\rm ring}$ extends PID to higher particle momenta, cf.
Fig.\,\ref{fig4.PID}. 
The effective space resolution of the system is about 5\,mm, which leads to $\sigma_{\rm ph}\approx 2.4$\,mrad. 
Finally, we expect 2$\sigma$  pion-to-kaon separation at 50\,GeV/c and efficient PID for small angles 
$\theta_{\rm Ch} > 20\,\rm{mrad}$. 
\begin{figure}[t]
\begin{minipage}[c]{.48\linewidth}
\begin{center}
\includegraphics[clip,bb= 14 20 398 342,width=1.\linewidth]{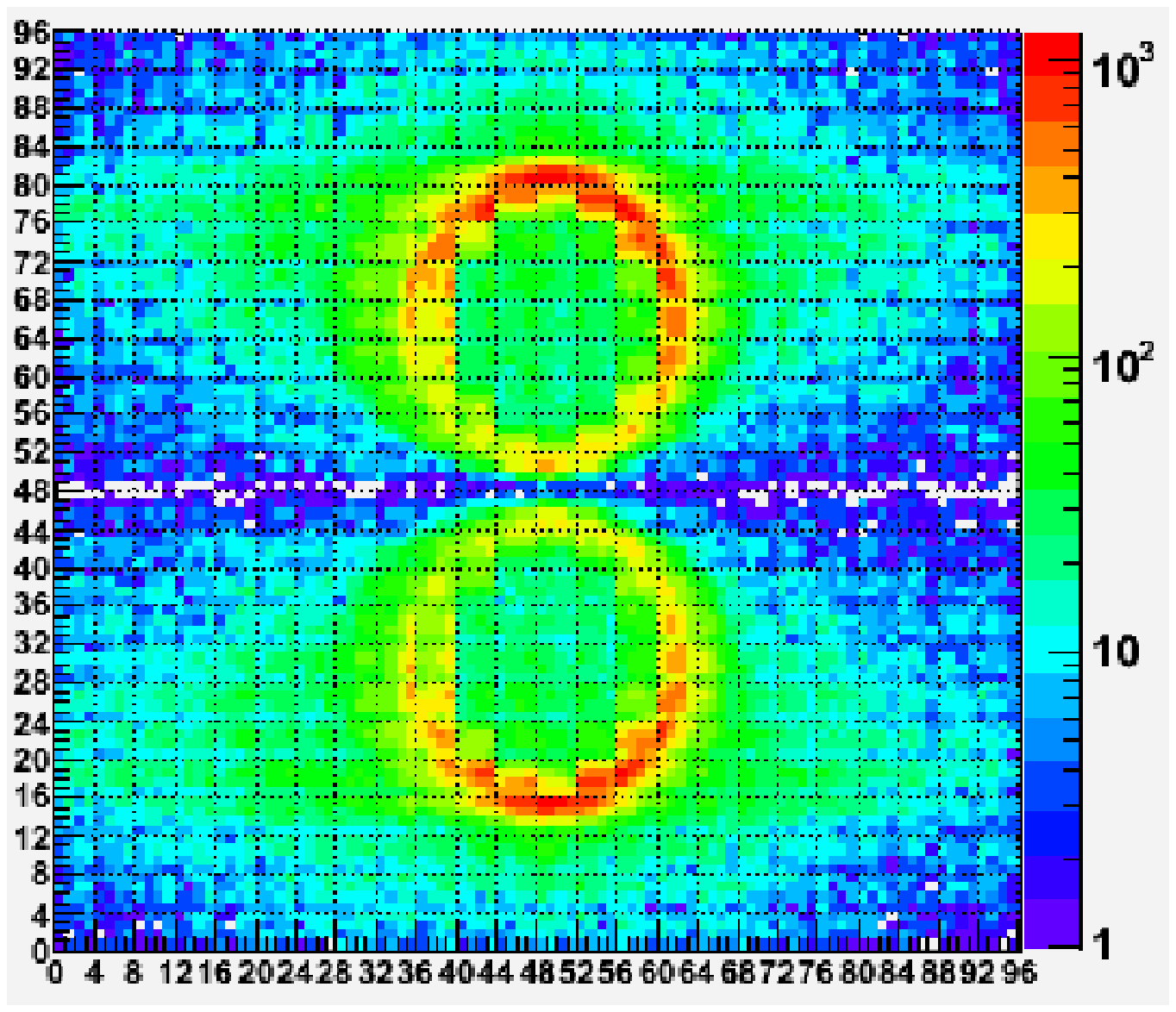}
\end{center}
\caption{Accumulated detector hitmap imaging the halo of beam particles.}
\label{fig5.coool_muon}
\end{minipage}\hfill
\begin{minipage}[c]{.48\linewidth}
\begin{center}
\includegraphics[clip,bb= 14 20 397 337,width=1.01\linewidth]{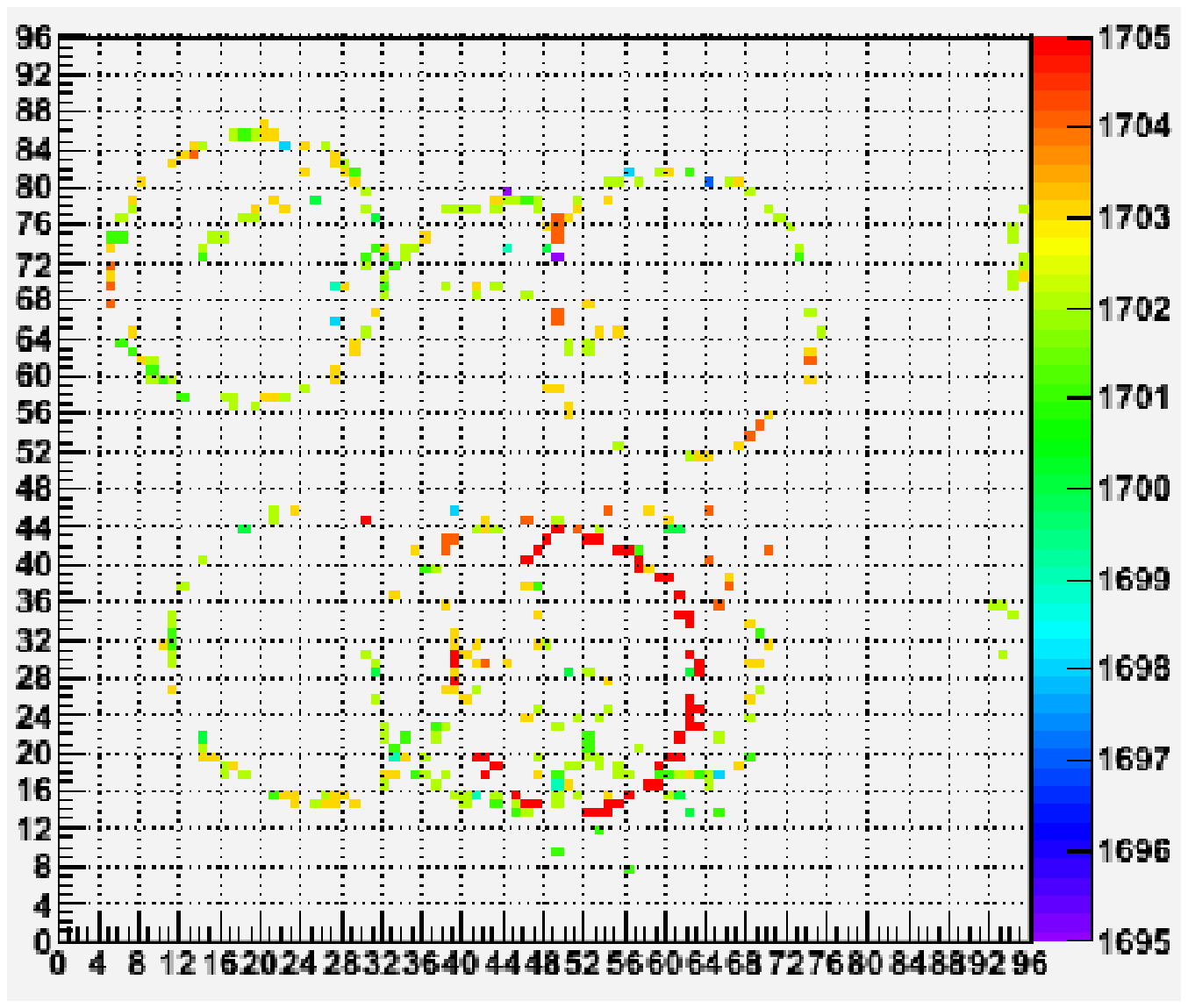}
\end{center}
\caption{Single physics event with multiple hadron rings (blue-red\,=\,10\,ns).}
\label{fig6.coool_singleevt}
\end{minipage}
\end{figure}
\newline
In Fig.\,\ref{fig5.coool_muon}, the two dimensional
hitmap of the $96\,\times\,96$ channels or pixels are shown for one of the first COMPASS 2006 SPS physics runs. One clearly
recognises the halo of the beam particles in the centre. 
First single event hadron rings detected with the COMPASS RICH MAPMT are shown in Fig.\,\ref{fig6.coool_singleevt} 
(10\,ns time cut applied). 
\vspace{-0.6 cm}
%
%
\section{Conclusions}
\label{conclusion}
\vspace{-0.8 cm}
For the upgrade of the COMPASS RICH-1, a fast photon detection system based on MAPMT was designed and
implemented on the time scale of one and a half years, and thus was ready for the COMPASS 2006 data taking. 
The PID capabality will be extended to both - high particle momenta and near the threshold. 
First signals in the detector indicate the new system is working well, improving future COMPASS physics results.
\vspace{-0.6 cm}
\section{Acknowledgements}
\label{acknow}
\vspace{-0.8 cm}
We acknowledge support from the CERN/PH groups PH/TA1,
TA2, DT2, TS/SU, and support by the BMBF (Germany) and the European Community-research
Infrastructure Activity under the FP6 programme (Hadron Physics, RII3-CT-2004-506078). 

\end{document}